# Widening of the Andes: an interplay between subduction dynamics and crustal wedge tectonics

Joseph Martinod[(1)], Mélanie Gérault[(2)], Laurent Husson[(3)], Vincent Regard [(4)]

(1) ISTerre, Université Grenoble Alpes, Université de Savoie Mont Blanc, CNRS, IRD, IFSTTAR, 38000 Grenoble, France. joseph.martinod@univ-smb.fr

(2) Department of Earth, Atmospheric and Planetary Sciences, Massachusetts Institute of Technology, Cambridge, MA 02139, U.S.A.

(3) Géosciences Environnement Toulouse, Université de Toulouse, CNRS, UPS(OMP), IRD, CNES, 14 av. E. Belin, 31400 Toulouse, France.

**Abstract:**

Shortening of the continental lithosphere is generally accommodated by the growth of crustal wedges building above megathrusts in the mantle lithosphere. We show that the locus of shortening in the western margin of South America has largely been controlled by the geometry of the slab. Numerical models confirm that horizontal subduction favors compression far from the trench, above the asthenospheric wedge and steeply dipping segment of the subducting slab. As a result, a second crustal wedge grows in the hinterland of the continent, and widens the Andes. In the Bolivian orocline, this wedge corresponds to the Eastern Cordillera, whose growth was triggered by a major episode of horizontal subduction. When the slab returned to a steeper dip angle, shortening and uplift pursued, facilitated by the structural and thermo-chemical alteration of the continental lithosphere. We review the successive episodes of horizontal subduction that have occurred beneath South America at different latitudes and show that they explain the diachronic widening of the Andes. We infer that the present-day segmented physiography of the Andes results from the latitudinally variable, transient interplay between slab dynamics and upper plate tectonics over the Cenozoic. We emphasize that slab flattening, or absence thereof, is a major driving mechanism that sets the width of the Andes, at any latitude.

**1. Introduction: modes of shortening of the continental lithosphere**

The Andean Cordillera is the longest, largest, and highest active margin orogen worldwide. It is located above a subduction zone that has remained active at least since the beginning of the Jurassic (Coira et al., 1982; Parada et al., 2007; Charrier et al., 2007; Mamani et al., 2010), although with possible interludes (Chen et al., 2019). However, Andean growth only began in the upper Cretaceous and is essentially younger than 50 My (Mégard, 1984; Oncken et al., 2006; Barnes et al., 2008; McQuarrie et al., 2008; Oncken et al., 2012; Armijo et al., 2015; deCelles et al., 2015; Faccenna et al., 2017; Parks & McQuarrie, 2019). The reasons that explain the spectacular growth of the Andes, quite sudden relative to the longer term subduction processes, are still debated (e.g., Chen et al., 2019). Moreover, it is now largely admitted that the growth of the chain has not been synchronous everywhere and that, for instance, pulses of shortening in some segments were coeval with extension in other parts of the Andes (Jordan et al., 2001; Mpodozis & Cornejo, 2012; Charrier et al., 2015; Horton & Fuentes, 2016; Horton, 2018b). Indeed, the Andes are highly segmented, some segments of the chain such as the Bolivian orocline being particularly high and wide, in contrast with others where the chain remains more modest both in elevation and width (Jordan et al., 1983; Mpodozis & Ramos, 1990; Kley et al., 1999; Lamb & Davis, 2003; Martinod et al., 2010; Horton, 2018a). The Andean segmentation is largely related to contrasted amounts of shortening of the South American continental plate. Although uncertainties remain on both the effect of magmatism on the generation of silicic crust (e.g., Ward et al., 2017) and on that of underplating resulting from the erosion of the continental plate by the subduction process (e.g., Baby et al., 1997), crustal shortening explains most of the present-day volume of the Andean crust and its latitudinal variations (Isacks, 1988; Kley & Monaldi, 1998; Kley et al., 1999; Arriagada et al., 2008; Eichelberger et al., 2015; Schepers et al., 2017). There is first-order geological evidence that the width of the Andes grew during discrete pulses (see Horton, 2018a, 2018b, for review). This discontinuous evolution is well documented in the Bolivian orocline, where the Eocene onset of shortening in the Eastern Cordillera resulted in a major and rapid eastward jump of the retro-arc foreland basin (e.g., deCelles & Horton, 2003).

The geological record provides reliable information on the temporal evolution of crustal deformation, but less so on the corresponding structures in the mantle lithosphere, where the style and geometry of deformations are likely decoupled from those that affect the upper crust. Geophysical data reach deeper levels of the lithosphere (e.g., Yuan et al., 2000; Oncken et al., 2003; Farias et al., 2010; Tassara & Echaurren, 2012; Chiarabba et al., 2016; Ryan et al., 2016; Delph et al., 2017; Ward et al., 2017), but only image the present-day structure. Although magmatic rocks constitute key data to understand the deep evolution of the Andes and its relationships with subduction dynamics (e.g., Soler et al., 1989; Soler & Bonhomme, 1990; Clark et al., 1990; Allmendinger et al., 1997; Kay & Mpodozis, 2002; Kay & Coira, 2009), mechanical models can provide further insights into the interplay between subduction dynamics and the tectonic evolution of the orogen.

In the following, we review existing models of the structural evolution of mountain ranges during continental shortening. They suggest that mountain ranges result from the growth of crustal wedges and accordingly, we show geological evidence that supports this concept for the Andean Cordillera. We argue that the widest parts of the range, however, do not correspond to a single expanding wedge in which the continuous propagation of new thrusts steadily broadens the external parts of the orogen. We propose instead that they result from the presence of two crustal wedges that are distinct laterally. We relate the development of a second crustal wedge to major tectonic reorganizations, controlled by the mechanical coupling between the slab and the continental mantle lithosphere, which in turn sets mechanical boundary conditions on the continental crust. We propose that the constantly evolving geometry of the slab largely explains the temporally and geographically contrasted evolution of the Andean Cordillera. We illustrate how slab geometry affects lithospheric stresses using mechanical finite-element models, and conclude that slab flattening is a primordial mechanism that sets the width of the Andes at any latitude.

**2. Shortening, structures and widening of mountain ranges: insights from mechanical models**

**2.1. Mantle lithosphere controls on crustal wedges**

In this section, we describe simple wedge models that have been used to analyze the shortening of the continental lithosphere. These models were used to explain the structure of collisional orogens. In the following sections, we show that because Andean growth essentially results from continental shortening, it shares many aspects with collisional orogens, and its evolution can largely be explained considering lithospheric scale wedge models.

The default model for collisional orogens depicts the convergence between two continental plates, following the subduction of the oceanic lithosphere that separated them. This situation is illustrated by the extensively studied Alpine orogeny (e.g., de Graciansky et al., 2011). Alpine-type collision chains grow in the aftermath of ocean closure, and it has been shown that their global structure can be compared with that of a crustal accretion prism developing above a mantle thrust (Beaumont et al., 1996; Escher & Beaumont, 1997; de Graciansky et al., 2011, Figure 1). They consist in a bivergent crustal wedge that grows as soon as the continental crust from the subducting plate can no longer be accommodated within the subduction channel. The growth of bivergent accretion prisms above deep megathrusts has been described using analogue model experiments for over 30 years (e.g., Malavieille et al., 1984; Colletta et al., 1991; McClay et al., 2004), and their mechanical behavior was further analyzed using numerical models in which thermal processes were accounted for (e.g., Beaumont et al., 1996; Vogt et al., 2017; Jaquet et al., 2018). These models show that further convergence results in increasing the width of the orogen, with in-sequence appearance of new thrusts in the external part of the crustal wedge, above the subducting mantle (Figure 1).

These simple analogue and numerical models that simulate the growth of an accretionary prism above a velocity discontinuity give a first order description of both sedimentary wedges in subduction zones, and many crustal-scale collisional wedges. The reason for this similarity is that similar boundary conditions apply to those systems. Their basal conditions can be modeled using two converging rigid plates, because in both cases, the basal layer is rigid compared to the wedge above it. Sediments deform more easily than the basement; in the same way, the continental crust generally requires lower deviatoric stresses to deform than the mantle lithosphere, as inferred from lithosphere rheological profiles (Chen & Molnar, 1983; Turcotte et al., 1984; Buck, 1991; Davy & Cobbold, 1991; Ranalli, 1995; Brun, 2002; Watts & Burov, 2003; Handy & Brun, 2004; Burov & Watts, 2006; Burov, 2011; Raimondo et al., 2014). Then, crustal wedges grow upwards and laterally outwards as convergence increases to accommodate the shortening imposed by the mantle lithosphere megathrusts. In numerical experiments that include thermal effects, crustal thickening is accompanied by heating of the lower crust, which weakens the lower part of the crustal wedge, limiting its elevation and favoring its widening instead (Figure 2a, Willett et al., 1993; Beaumont et al., 2004; Gerbault & Willingshofer, 2004).

In contrast, if the mantle lithosphere is weak, the boundary conditions that apply to the base of the crust do not correspond anymore to those applied in models of accretionary prisms. Models confirm that the geometry of crustal structures would largely differ from crustal wedges developing above mantle megathrusts in the absence of any strong mantle lithosphere (Davy & Cobbold, 1991; Cagnard et al., 2006a, 2006b).

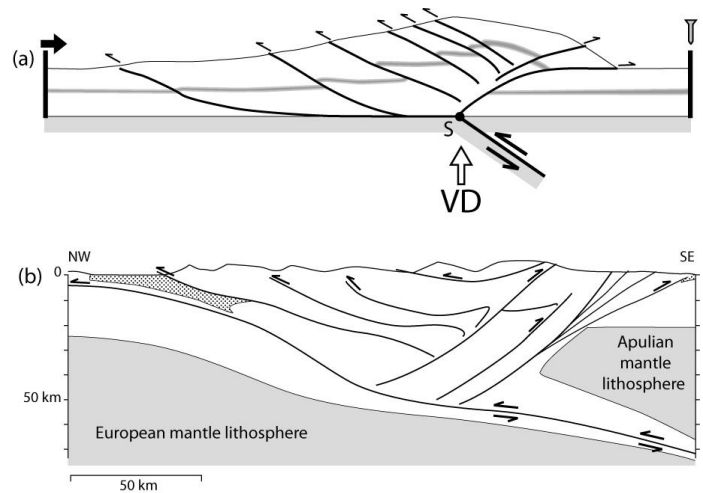

*Figure 1: (a) growth of a brittle accretionary wedge above a basal discontinuity (after Malavieille, 1984). VD is the basal velocity discontinuity. A single back-thrust forms at the beginning of shortening, and the progressive in sequence appearance of pro-thrusts above the subducted plate results in the widening of the wedge. (b) Structure of the Western Alps, after Escher & Beaumont (1997). The alpine range geometry is that of a double-vergent crustal wedge, which shares many similarities with the geometries obtained in the simple model presented above, when considering that the base of experiments represents the rigid mantle lithosphere.*

## 2.2 Widening the range: single or multiple crustal wedges?

If the mantle lithosphere is sufficiently rigid, models show that a single crustal wedge progressively widens to accommodate the convergence between lithospheres. This process has been proposed to persist as long as convergence continues, and to eventually result in the formation of crustal wedges whose width may be as large as that of the Tibetan Plateau (Beaumont et al., 2004). According to Beaumont et al. (2004), plateau widening may essentially result from the activity of a unique velocity discontinuity ("S detachment point") within the mantle lithosphere. Comparable models of outward growth of a single crustal wedge have been proposed to explain the widening of the Central Andes (e.g., Isacks, 1988; Allmendinger & Gubbels, 1996; Allmendinger et al., 1997; Pope & Willett, 1998; Husson & Ricard, 2004). However, the geological record is at odds with this concept of continuous expansion. Widening was instead marked by a major shortening gap in the Altiplano, accompanied by a consistent jump of tectonic shortening towards the Eastern Cordillera in the Eocene (see below and McQuarrie & deCelles, 2001; deCelles & Horton, 2003; Elger et al., 2005; Oncken et al., 2006; Anderson et al., 2018; Horton, 2018a). Despite this evidence, models of the Central Andes often depict a single crustal wedge wherein shortening in the mantle lithosphere concentrates beneath the Western Andes, and wherein a long and flat intra crustal thrust transfers surface shortening towards the Eastern Cordillera, Interandean and Subandean zones (e.g., Isacks, 1988; McQuarrie & deCelles, 2001; deCelles & Horton, 2003; McQuarrie et al., 2008; Eichelberger et al., 2015; Anderson et al., 2017; Anderson et al., 2018).

Conversely, mechanical models simulating the shortening of the continental lithosphere show that orogen widening may also result from the activation of multiple thrust faults crossing the mantle lithosphere. This is particularly true if the mantle is not very rigid and/or if it consists of accreted blocks separated by ancient suture zones (Martinod & Davy, 1994; Burg et al., 1994; Sobolev & Babeyko, 2005; Jourdon et al., 2017, Figure 2). The case of Tibet, which results from the accretion of different continental blocks (e.g., Allègre et al., 1984), can provide some insights. Models that challenge the continuous plateau evolution proposed by Beaumont et al. (2004) describe the thickening of the Tibetan crust and the widening of the plateau as a result of the piecemeal activation of several faults in the mantle lithosphere (Mattauer, 1986; Burg et al., 1994; Tapponnier et al., 2001). The crustal structures and the evolution of the orogen largely differ depending on strain localization in the mantle. If several mantle faults are activated, the orogen consists of several coalescing crustal wedges (Martinod & Davy, 1994; Burg et al., 1994; Tapponnier et al., 2001). Compressionnal basins, limited on both sides by reversed faults of opposite vergence, appear between adjacent wedges. In the following, we discuss the structural evolution of the Central Andes in light of these models. We argue that the evolution of the Central Andes does not correspond to the growth of a single widening crustal wedge but instead to the growth of distinct major wedges overlying different mantle thrusts, whose activity is controlled by the geometry of the slab. We show that lithospheric scale structures accommodating the growth of the high plateau in the Central Andes share many common aspects with those that have been proposed in the models cited above that describe the widening of the Tibetan Plateau.

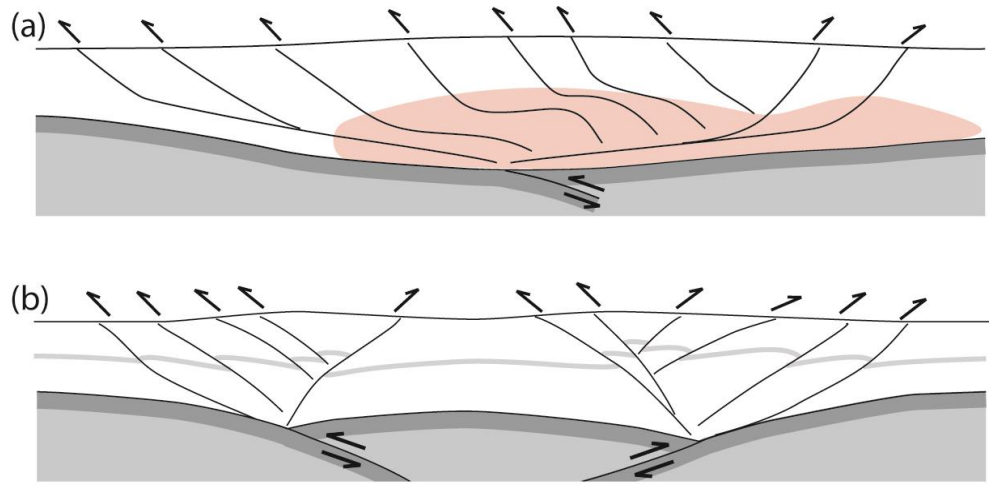

***Figure 2:*** *Two possible mechanisms that result in the widening of a continental collision zone. (a) Widening of a crustal wedge above a single mantle megathrust (redrawn from models presented in Beaumont et al., 2004, without any surface denudation). The red area corresponds to the basis of the wedge above 700°C. (b) Widening of the collision zone through successive mantle thrusts over which crustal wedges grow (redrawn from models presented in Martinod & Davy, 1994)*

## 3. Tectonic evolution of the Central Andes

The Andes are the widest and highest in their central segment between 5°S and 33°S, where continental shortening was the most intense (Kley & Monaldi, 1998; Kley et al., 1999; Arriagada et al., 2008; Eichelberger et al., 2015; Faccenna et al., 2017; Garzione et al., 2017; Schepers et al., 2017; Horton, 2018b). Andean segmentation, i.e. the contrast existing between segments of the range at different latitudes, is particularly marked at the boundary between the Central and the Southern Andes (e.g., Giambiagi et al., 2012). The Southern Andes, indeed, correspond to a narrow chain, at lower average elevation, and have accommodated significantly smaller amounts of crustal shortening.

The growth of the Central Andes initiated in the Upper Cretaceous (Peruvian phase of Steinman, 1929), possibly resulting from the westward drift of South America following its separation from Africa (e.g., Frutos, 1981; Dalziel, 1986; Silver et al., 1998; Somoza & Zaffarana, 2008; Ramos, 2010; Husson et al., 2012). During Jurassic and Lower Cretaceous, the tectonic regime of the South American active margin was generally under extension (e.g., Dalziel, 1981; Mpodozis & Ramos, 1990; Jaillard et al., 1990; Coney & Evenchick, 1994; Charrier et al., 2007; Ramos, 2010; Horton, 2018a). Apart from volcanoes, mountain ranges were absent from the margin and the magmatic arc was generally separated from the onshore continent (Brazilian craton) by shallow epicontinental seas (Aubouin et al., 1973; Mpodozis & Ramos, 1990; Jaillard & Soler, 1996; Ramos, 2010). At that time, even though subduction had been active for at least 100 My (Early Jurassic, Coira et al., 1982; Mamani et al., 2010; Oliveros et al., 2006), no relief had developed along the western boundary of South America and the physiography of the margin was somewhat comparable to that of the present-day Indonesian arc.

Cretaceous shortening initiated together with a first pulse of hinterland migration of the magmatic arc, probably reflecting the diminution of the slab dip angle (Figure 3, Mégard et al., 1984; Scheuber & Reutter, 1992; Jaillard et al., 2000; Haschke & Gunther, 2003; Mamani et al., 2010; Ramos, 2010; Charrier et al., 2015; Gianni et al., 2018). During Upper Cretaceous and Paleocene, basin analysis shows that the chain remained limited to the Precordillera, west of the present-day volcanic arc at the latitude of the future Bolivian Orocline (Horton et al., 2001; deCelles & Horton, 2003), as well as in the southern Central Andes (Mpodozis et al., 2005; Arriagada et al., 2006; Bascuñan et al., 2016, 2019; deCelles et al., 2015; Henriquez et al., 2019). It implies that the Central Andes were then narrow, probably comparable in width to the present-day Southern Andes. The foreland retro-arc basin, where clastic sediments accumulated, was limited to the Western Altiplano during upper Cretaceous. Sediments deposited at the locus of the future Eastern Cordillera accumulated in a back-bulge region, far from the range (deCelles & Horton, 2003). Sediment provenances in these series confirm this back-bulge context, as they show cratonic sources and no influx from the magmatic arc (Horton, 2018a).

At the latitude of the Bolivian orocline, the growth of the Eastern Cordillera started during the Eocene (Mégard et al., 1987; Benjamin et al., 1987; Lamb & Hoke, 1997; Lamb et al., 1997; McQuarrie & deCelles, 2001; Horton, 2005; Ege et al., 2007; Garzione et al., 2017; Rak et al., 2017; Anderson et al., 2018). Shortening began by activating west verging thrusts in the western part of the Eastern Cordillera (Hérail et al., 1993; Kley et al., 1997; Lamb & Hoke, 1997; Sempere et al., 1999; McQuarrie & deCelles, 2001; Muller et al., 2002; McQuarrie, 2002a; McQuarrie et al., 2008; Eichelberger et al., 2013; Perez et al., 2016; Anderson et al., 2017, 2018) and east verging thrusts in the eastern part of the range. This period resulted in rapid subsidence and accumulation of terrigenous sediments coming from the East, in what is today the western margin of the Eastern Cordillera (Lamb et al., 1997). This reveals that the proto-Eastern Cordillera had developed, isolating a large hinterland basin roughly corresponding to the future Altiplano, from a new eastern foreland basin (Horton & deCelles, 2003; McQuarrie et al., 2005; Horton, 2018a). On the western side of this newly formed Eastern Cordillera, compressive structures progressed westward, from the central part of the Eastern Cordillera towards the future Altiplano (Murray et al., 2010; Anderson et al., 2018). It implies that the widening of the Central Andes in Bolivia did not result from the monotonous eastward progradation of east-verging thrust faults. In other words, the Central Andes cannot be described as a single crustal wedge monotonically widening from its initial position adjacent to the trench, but by the inception of discrete events across-strike.

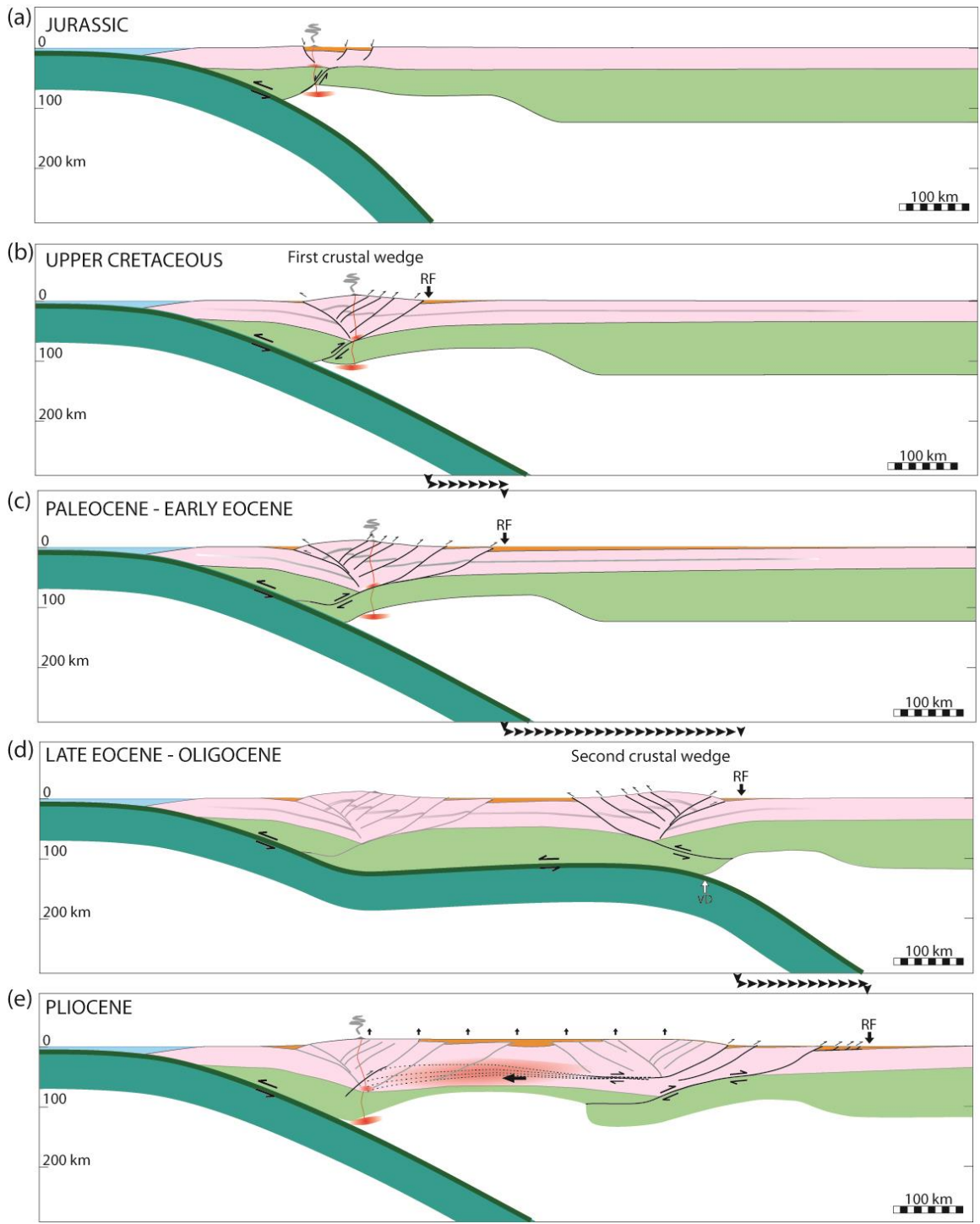

***Figure 3:*** *Evolution of the Andes across the Bolivian orocline (in the reference frame of the trench). (a) Jurassic and Lower Cretaceous, the tectonic context of the margin is extensive above a steep oceanic slab; (b) Upper Cretaceous: beginning of continental shortening, resulting in the growth of the Western Cordillera above a gently-dipping slab segment; (c) continental shortening results in the growth of a crustal wedge above a deep mantle thrust; (d) Late Eocene: the appearance of a horizontal slab segment increases continental shortening. A new crustal wedge appears in the Eastern Cordillera above a new mantle thrust zone; (e) Pliocene: oceanic slab is inclined again beneath this part of the Andes. The hot, hydrated, and thickened crustal root of the chain facilitated deep crustal flow accommodating part of the uplift of the Altiplano. The topography of the plateau opposes shortening, which is transferred east of the plateau in the Subandean Ranges. RF marks the position of the Retro-arc foreland Basin and the arrows between successive images show its eastward migration.*

West-verging structures located East of the Altiplano are commonly described as back-thrusts that initiate at the eastern termination of a flat east-verging middle-lower crust fault (Roeder & Chamberlain, 1995; Baby et al., 1997; McQuarrie & deCelles, 2001; Horton, 2005; Barnes et al., 2008; Espurt et al. 2011; Eichelberger etal., 2013, 2015; Anderson et al., 2017). This flat fault would transfer shortening accommodated in the mantle lithosphere beneath the Western Cordillera towards the Eastern Cordillera. Although geometrically consistent, this mechanism requires a major strength contrast between the upper and middle crust to generate the flat structure transferring the shortening within the lower crust on hundreds of kilometers without any significant surface deformation. This may occur if the lower-middle crust viscosity is particularly weak, which allows to largely decouple surface deformations from mantle shortening (Royden et al., 1997; Vanderhaeghe et al., 2003; Gerbault & Willingshofer, 2004; Flesch et al., 2005; Chardon et al., 2011; Chen & Gerya, 2016). This mechanism may thus occur in high plateaus such as Tibet and Altiplano, in which the thickness and temperature profile of the crust result in particularly small middle crust viscosity (Royden et al., 1997; Beaumont et al., 2001; Husson & Semperé, 2003). In contrast, the rheological profile of the continental lithosphere in the future Altiplano, before tectonic shortening, may not have permitted to fully decouple the mantle from surface deformations.

The growth and widening history of the Andes can also be unraveled from the sediment record in foreland retro-arc basins, provided that dynamic topography does not blur the signal (e.g., Davila and Lithgow-Bertelloni, 2013; Flament et al., 2015). In the Bolivian oroclinef, Paleogene sediments record the rapid migration of shortening towards the Eastern Cordillera. It was accompanied by a rapid eastward migration of the retro-arc foreland basin, contrasting with the modest eastward migration that prevailed during Upper Cretaceous, and faster than the eastward shift that occurred afterwards during the Neogene (Figure 3, deCelles & Horton, 2003).

The development of the Eastern Cordillera was coeval with a major perturbation of the magmatic arc in the Central Andes. The volcanic arc remained relatively stable between the beginning of the Upper Cretaceous and the Paleocene; then, 45 My ago, it moved rapidly more than 200 km north-eastward in South Central Peru (see Figure 4 and Mamani et al., 2010). In southern Peru and northern Chile, the Eocene is marked by the extinction of the volcanic arc. At present-day, horizontal slab segments beneath South America correspond to a lull in volcanic activity. As the slab flattens and comes in contact with the overriding plate, it blocks the inflow of hot asthenosphere into the wedge; volcanism first migrates inboard from the trench, and eventually stops (e.g., McGeary et al., 1985; Gutscher et al., 2000a; English et al., 2003). Hence, the magmatic evolution of the Central Andes 45 My ago has been attributed to the inception of a horizontal slab segment (e.g., Soler & Bonhomme, 1990; Sandeman et al., 1995; James & Sacks, 1999; Kay et al., 1999; Ramos & Folguera, 2009; Martinod et al., 2010; O'Driscoll et al., 2012).

The end of flat-slab subduction and the return to a steeply dipping slab during the Oligocene and the Miocene resulted in widespread melting of the lithosphere of both the Eastern Cordillera and the Altiplano, consequence of the influx of hot asthenosphere beneath a hydrated continental plate (James & Sacks, 1999). Voluminous ignimbritic volcanism also affected the Western Cordillera, first in Southern Peru at the end of the Oligocene, and then in Northern Chile in the Lower Miocene. The return to a steeply dipping slab beneath the Central Andes was also marked by a westward migration of shortening from the Eastern Cordillera towards the Altiplano (McQuarrie et al., 2005; Anderson et al., 2018). Although Miocene thrust faults also deformed the Western Cordillera (Muñoz & Sepulveda, 1992; Muñoz & Charrier, 1996; Victor et al., 2004; Garcia & Hérail, 2005; Farias et al., 2005; Hoke et al., 2007; Charrier et al., 2013; van Zalinge et al., 2017), they are steep and account for very small amounts of crustal shortening. Victor et al. (2004) estimate that only 2.5 km of Neogene shortening occurred in the Western Cordillera. If so, Miocene thrust faults in the Western Cordillera may only accommodate the relative uplift of Altiplano with respect to the forearc region. They do not explain crustal thickening beneath the Central Andes. Part of the uplift has also been accommodated by the westward tilting of the western flank of the Altiplano. It shows that uplift occurred independently of any superficial tectonic shortening in this sector, and resulted from lower crustal flow (Isacks, 1988; Schildgen et al., 2007; Thouret et al., 2007; Jordan et al., 2010).

Tectonic shortening occurred in the Altiplano during the Miocene, but largely decreased since Late Miocene (Hérail et al., 1996; Lamb & Hoke, 1997; Lamb, 2011), although minor faulting has occurred as recently as Pliocene (Lavenu & Mercier, 1991). In fact, most of the Neogene continental shortening in the Central Andes occurred in the Eastern

Cordillera and in the Subandean ranges (Dunn et al., 1995; Kley, 1996; Baby et al., 1997; McQuarrie, 2002a; Muller et al., 2002; McQuarrie et al., 2008; Gotberg et al., 2010; Eichelberger et al., 2013; Perez et al., 2016; Anderson et al., 2017). The Neogene also corresponds to a major episode of uplift of the plateau (Garzione et al., 2017), as suggested by oxygen isotopes analyses (Garzione et al., 2006, 2014), paleobotanic data (Gregory Wodzicky, 2000; Kar et al., 2016; Perez-Escobar et al., 2017), and geological observations on the western flank of the Andes (Lamb et al., 1997; Farias et al., 2005; Garcia & Hérail, 2005; Jordan et al., 2010).

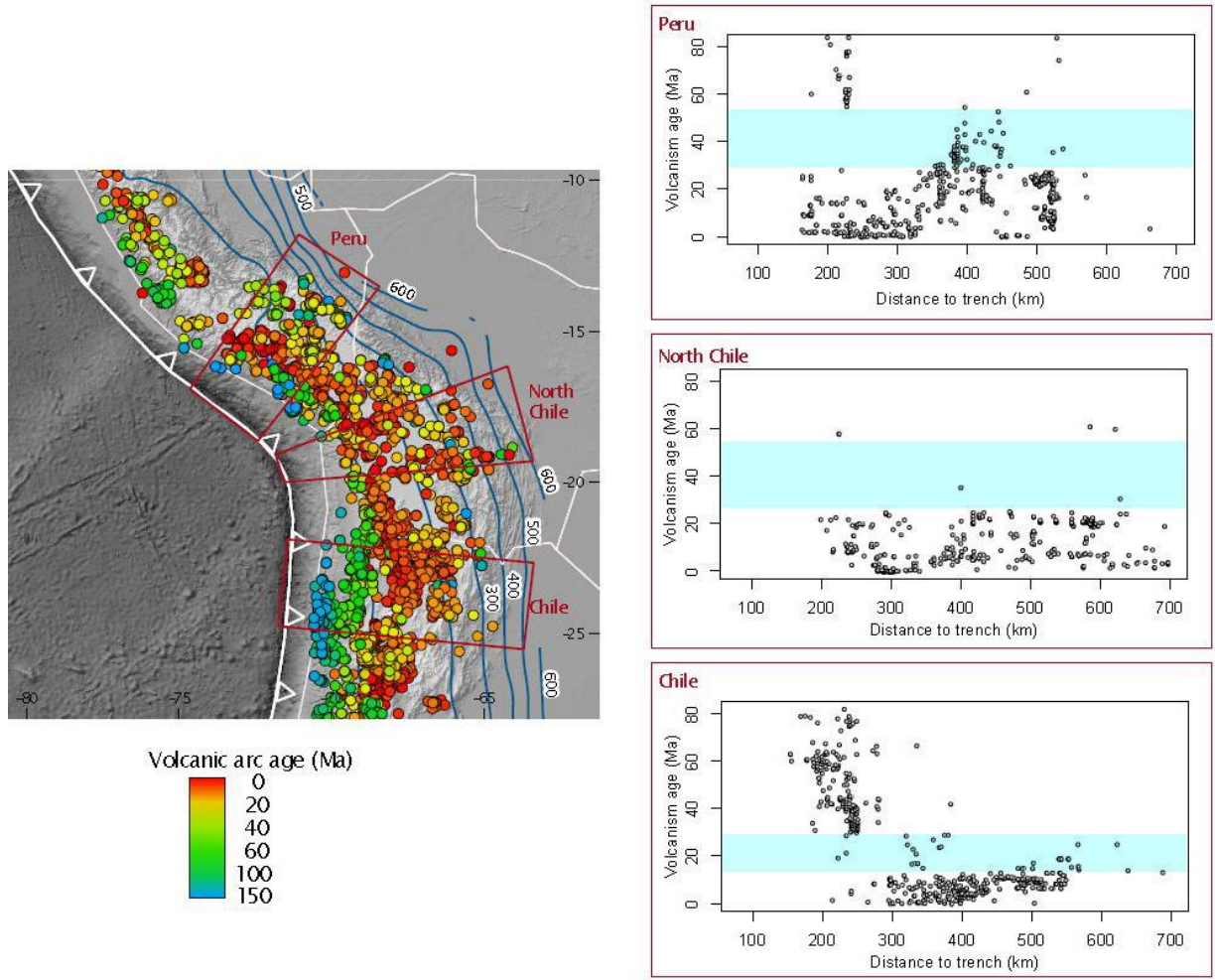

*Figure 4: Age of volcanism in the Central Andes vs. distance to the trench, along 3 transects in Southern Peru, in northern Chile/Bolivia (Bolivian orocline), and in Chile/Argentina between 22° and 25°S (Puna). Data from Mamani et al. (2010) and the Central Andes geochemical GPS database web site (https://andes.gzg.geo.uni-goettingen.de/). Blue segments mark periods during which the volcanism migrated towards the hinterland and eventually stopped, due to horizontal subduction (after James & Sacks, 1999, for the Southern Peru and Bolivian orocline; Kay et al., 1999, for the Puna Plateau).*

## 4. Horizontal subduction: a key to understanding the Paleogene evolution of the Cordillera

In this section, we describe how the Late Eocene-Oligocene flat-slab subduction beneath the Central Andes may have controlled both the along-strike and across-strike evolution of the orogen. We also suggest that the subsequent Miocene uplift of the Altiplano region is largely inherited from this episode of horizontal subduction.

## 4.1. Along-strike effect of the Paleogene flat-slab subduction in the Central Andes

Sediment accumulation and structural markers indicate that the Paleogene evolution of the Bolivian orocline contrasts markedly with that of other segments of the Andes. For instance, Horton (2018a) describes the eastward migration of sedimentary basins accumulating materials eroded from the Cordilleran range. He observes a specific period in the Paleogene during which sediment accumulation rates severely decreased everywhere. He proposes that it may reflect an episode of moderate shortening (or extension), in contrast to the significant shortening that prevailed before and after this period. In the Southern Andes, this period is marked by extension, with the appearance of grabens in which the sediments of the Abanico formation deposited (e.g., Charrier et al., 2002; Folguera et al., 2002; Orts et al., 2012). In the Colombian Andes, the Paleogene is also marked by a major sedimentary hiatus accompanying a cessation of shortening (Gomez et al., 2003; 2005). Although a decrease in shortening rates and a diminution in sediment accumulation rates were indeed observed in many segments of the Andes, the evolution of the Central Andes during this time was markedly different, especially in the Bolivian orocline.

This is also evidenced by paleomagnetic data obtained in the forearc area of the Andes (Roperch et al., 2006), which indicated that both clockwise rotations in Northern Chile and counter-clockwise rotations in Southern Peru mainly occurred over this rather limited time period (Late Eocene – Oligocene). These rotations recorded the indentation of the active margin of South America that resulted in the major episode of shortening that created the Bolivian orocline (Figure 5, Arriagada et al., 2008). Martinod et al. (2010) proposed that this pulse of shortening in the Central Andes resulted from horizontal subduction, triggered by the subduction of an oceanic plateau. Horizontal subduction increases interplate coupling, which may explain larger amounts of shortening transmitted to the upper plate and a decrease in subduction rates. It is compatible with plate indentation by an aseismic ridge and the associated tectonic rotations.

We propose that the segmentation of the Andean Cordillera was triggered by episodic horizontal slab segments beneath South America (see also e.g., Jordan et al., 1983; Jordan & Allmendinger, 1986; Ramos, 2009; Martinod et al., 2010). Lithospheric-scale models suggest that a local perturbation such as, for instance, the subduction of a plateau or a ridge, may affect the overriding plate tectonic regime above the perturbed area, but also the entire subduction zone if the perturbation is large enough to alter the dynamics of the whole subduction system (e.g., Regard et al., 2005; Faccenna et al., 2006; O'Driscoll et al., 2012; Arrial & Billen, 2013; Martinod et al., 2013; Florez Rodriguez et al., 2019). Mechanical models suggest that smaller convergence velocities favor extension in the overriding plate if other parameters, such as the negative buoyancy of the slab, upper plate absolute motion in the lower mantle reference frame, trench length, or interplate friction, remain unchanged (Schellart, 2005; Heuret et al., 2007; Guillaume et al., 2018; Cerpa et al., 2018). They also confirm that shortening in the overriding plate above the ridge may be accompanied by a decrease in subduction velocity, and by extension in other segments of the active margin (Martinod et al., 2013; Florez Rodriguez et al., 2019).

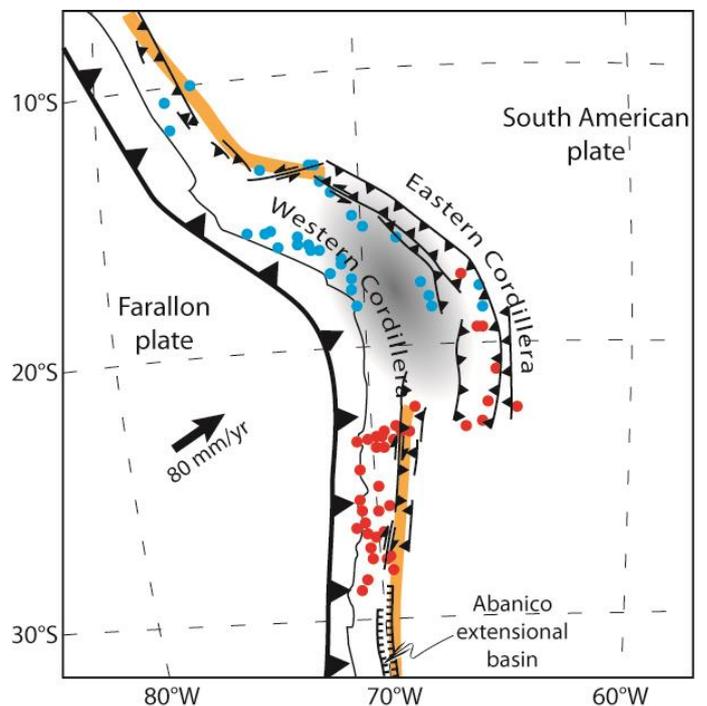

*Figure 5: Oligocene growth of the Bolivian orocline. Paleogene tectonic rotations are marked by red and blue dots (red = clockwise; blue = counterclockwise) after Arriagada et al. (2008). Oligocene structures after Mégard (1984), Charrier et al. (2015), Lamb (2001), Mpodozis & Cornejo (2012). The orange thick line corresponds to the magmatic arc, and the gray area marks the position of the Oligocene flat slab segment.*

Most plate kinematic models indicate a Late Eocene-Oligocene period of reduced convergence rates between the Farallon/Nazca oceanic plate and South America (Pilger, 1984; Pardo Casas & Molnar, 1987; Somoza, 1998; Sdrolias & Muller, 2006; Somoza & Ghidella, 2012), although an alternative model in which the convergence velocity increased before the Oligocene has been proposed by Wright et al. (2016). In the other scenarios cited above, the Oligocene convergence velocity is roughly two times smaller than the Miocene Nazca-South America convergence velocity. If an increase in convergence velocity at the Oligocene-Miocene boundary may be explained by the coeval fragmentation of the Farallon plate (Lonsdale, 2005), its Eocene diminution does not result from any major change in the geometry of the tectonic plates. We propose that the coupling in the horizontal slab segment underneath the Central Andes may have been increased enough to slow down the subduction rate of the Farallon Plate beneath South America and thus, may have favored continental extension southward. It follows that the impact of flat-slab subduction underneath the Central Andes would reach beyond the sole region of coupling between the flat slab and the overriding plate, but extend to the entire plate system. In other parts of the active margin, the decreased convergence velocity may have triggered extension (as for instance in the Abanico Basin close to Santiago de Chile) or at least, contributed to the widespread decrease of continental shortening rates reported by Horton (2018a). We thus envision a genetic link between the growth of the Eastern Cordillera in the Bolivian orocline, and extension observed in the Southern Andes. In section 5.5, we further discuss other mechanisms that may have contributed to both stronger compression in the Bolivian Orocline, and extension in the Southern Andes.

**4.2. Across-strike effect of flat-slab subduction in the Central Andes**

**4.2.1 Stresses in the overriding plate: insights from numerical models**

Horizontal subduction, by increasing the interplate contact area, promotes continental shortening and the hinterland migration of deformation (e.g., Bird, 1984; Gutscher et al., 2000a; Humphreys and Coblentz, 2007; Axen et al., 2018). The horizontal force that accumulates along a vertical section of the overriding plate progressively increases eastward, because of the basal friction exerted by the subducting plate on the continent (e.g., Dalmayrac & Molnar, 1981; Bird, 1984; Espurt et al., 2008; Martinod et al., 2010; O'Driscoll et al., 2012). Therefore, slab flattening offsets the maximum compressive stresses eastward. This effect is amplified by the negative buoyancy of the steep segment of the slab, which is shifted toward the continental interior during flat-slab subduction; thus, so are the compressive stresses that it induces.

The results of numerical models presented in Figure 6 illustrate this effect. We use a 2-D, cylindrical finite-element code adapted from MILAMIN (Dabrowski et al., 2008; Gérault et al., 2012), which has been used previously to study regional flat slab dynamics (Gérault et al., 2015; Siravo et al., 2019). The computation solves for the velocity and pressure fields using an infinite Prandtl number and incompressible Stokes flow formulation with Newtonian rheology. The models are instantaneous, and the density and viscosity fields are uniform within each prescribed structural domain (see supplementary material). The upper and lower boundary conditions are set to free slip. The cylinder is truncated on each side, where the boundary conditions are also free slip. These instantaneous models, despite being dynamic, do not address the time-evolution of the subduction system (for instance here, how the slab becomes flat). Our goal here is to assess stresses in the overriding plate to constrain the locus of shortening depending on the geometry of the slab and on the crustal and lithosphere thicknesses, making these instantaneous models suitable for our purpose. Recent time evolving flat slab models by two independent groups, Axen et al. (2018) and Sandiford et al. (2019), confirm that the stress patterns in the slabs are extremely similar in instantaneous (Gérault et al., 2015) and time-evolving models. Further details about the models (mesh, density and viscosity fields, geometrical parameters, etc.) are included in the supplementary material.

Figure 6 shows the stresses in the overriding lithosphere for a series of slab geometries, that represent different stages of Andean development (not strictly identical to the temporal stages depicted in Figure 3). All parameters are kept identical between the models, besides the geometry of the slab and the thickness of the continental crust and lithosphere. These models exemplify how a change in slab geometry drastically affects both the magnitude and the distribution of the effective stress in the overriding plate. During the initial stage (Figure 5a) the steep subducting slab causes minor stresses in the lithosphere. In practice, this low-stress regime is akin to the Jurassic setting of South America, at a time where no orogen had developed yet. When the slab is less steep (Figure 5b), compression prevails in the upper plate above the asthenosphere corner (volcanic arc). This situation could correspond to the Upper Cretaceous setting of the Andes, during the onset of crustal thickening. If the crust thickens during that stage (Figure 5c), buoyancy forces compensate the stresses that arise from convergence, and the tectonic regime becomes almost neutral (here, slightly extensive) in the thickened sections of the upper crust. When the slab flattens, compression reaches very high values above the hinge of the slab, at the location of the Eastern Cordillera, while only minor deformation occurs in the Altiplano (Figure 5d). In practice, it corresponds to a discrete eastward offset of the locus of maximal compression that bypasses the Altiplano. That would correspond to the Late Eocene stage in the Bolivian orocline. While compression proceeds in the entire range and in the Eastern Cordillera in particular (Oligocene, Figure 5e), compression is hindered in the Eastern Cordillera by the crustal buoyancy forces, but only to a minor extent. Last, when the slab returns to a steeper dip (Figure 5f), compression decreases and only remains active east of the uplifted region whose elevation is supported both by the thick crust and thin mantle lithosphere. This situation resembles the Central Andes at present-day. At all stages, models predict crustal extension close to the trench, which indeed is observed in the forearc region (e.g., Delouis et al., 1998; Allmendinger & Gonzalez, 2010; Farias et al., 2011; Comte et al., 2019). However, the forearc being cold and rigid, strain rates are smaller than elsewhere in the Andes and deformations that occur in this area largely decoupled from deformations that occur in the arc and back-arc region.

These models are only valid at first order, as a result of the assumptions discussed above. Yet, they yield the most important result of our study: flat slab subduction increases compression of the continental plate, and results in the eastward migration of the shortening. Compression concentrates above the asthenospheric corner, triggering the growth of a second crustal wedge. In the Bolivian orocline, the intense Eocene pulse of shortening, from which the Eastern Cordillera arises, directly results from the development of the flat slab.

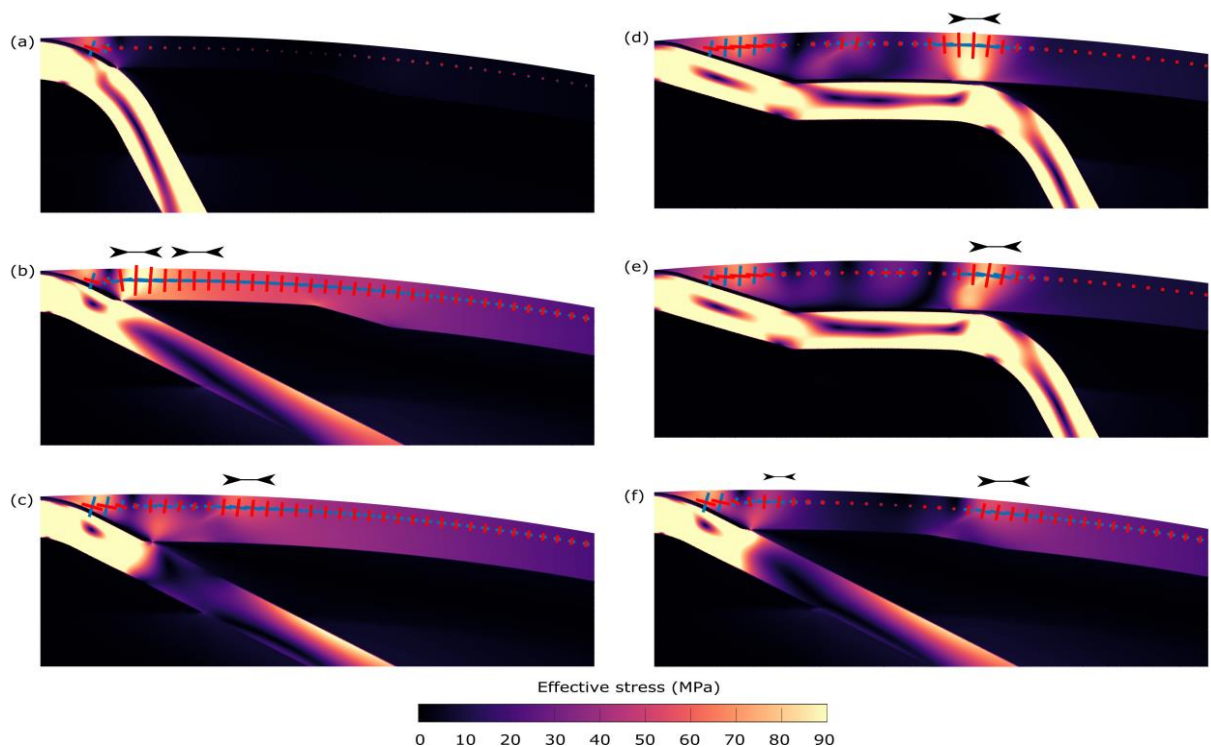

*Figure 6:* Numerical model predictions of the deviatoric stresses in the overriding plate above variable slab geometries. a) 60° slab dip angle; b) 30° slab dip angle before crustal thickening; c) 30° slab dip angle following crustal thickening in the western part of the continent; d) flat slab over 600 km from the trench, thicker continental crust on the western side; e) flat slab over 600 km from the trench, with a thicker continental crust from the trench to the easternmost extent of the flat slab; f) 30° slab dip angle, with a thicker continental crust from the trench to the easternmost extent of the flat slab. All parameters are kept identical between the models, besides the geometry of the slab and the thickness of the continental crust and lithosphere. The colored background shows the effective stress (square root of the second invariant of the deviatoric stress tensor). Axes and amplitudes of the principal deviatoric stress tensor mid-overriding plate are shown in blue (compression) and red (extension). See supplementary material for details.

### 4.2.2 Effect of slab flattening on the continental shortening accommodation by crustal wedges

Slab flattening is a relatively rapid phenomenon in the history of the Andean orogen (Soler & Bonhomme, 1990; Yañez et al., 2001; Gutscher et al., 2000b; Ramos et al., 2002; Rosenbaum et al., 2005; Espurt et al., 2008; Maksaev et al., 2009). Observations from volcanic arc migration suggest that it takes some 7 My to get a ~700 km wide (from the trench to the asthenospheric wedge) horizontal subduction zone (Gutscher et al., 2000b; Kay & Mpodozis, 2002; Ramos et al., 2002; Rosenbaum et al., 2005; Espurt et al., 2007; see discussion in Espurt et al., 2008). If the distance between the trench and the volcanic arc is ~300 km before slab flattening, the average migration rate of the mantle wedge beneath the continental plate is ~60 mm/yr, largely exceeding upper plate shortening rates. Mechanical models predict that as the locus of maximum deformation in the mantle lithosphere migrates eastward with the development of the flat slab (Figure 6), crustal deformation follows, as it is driven by the underlying stresses from the lithosphere. Tectonic activity ceases in the western crustal wedge, where lithospheric stresses have decreased. A new wedge appears inland, following the displacement of both the velocity discontinuity at the base of the crust and the steeply dipping segment of the subducting slab (Figure 3).

In the Central Andes in the Eocene, this eastward offset of shortening isolated a hinterland compressive basin (the Altiplano Basin), located between the Western and Eastern Cordilleras, from the foreland basin located East of the new crustal wedge (e.g., deCelles & Horton, 2003; McQuarrie et al., 2005; Horton, 2005). In the central and western part of the Eastern Cordillera, West verging thrust faults confirm that the growth of the range resulted from the initiation of a new crustal wedge, regardless of prior tectonic activity in the Western Cordillera. These thrusts sequentially propagated to the West in the Eastern Cordillera (Lamb & Hoke, 1997; McQuarrie & deCelles, 2001; Muller et al., 2002; Murray et al., 2010; Eichelberger et al., 2013; Anderson et al., 2018). If instead the Eastern Cordillera had simply resulted from the progressive widening of the western wedge, deformation would have regularly progressed eastward over east-verging thrusts.

Earlier studies proposed that west-verging thrusts in the Eastern Cordillera correspond to back-thrusts within a general landward transport to the East, along a flat structure located in the crust beneath the future Altiplano (Baby et al., 1997; McQuarrie & deCelles, 2001; Eichelberger et al., 2013, 2015; Anderson et al., 2017). This deformation pattern implies that shortening in the upper crust of the Eastern Cordillera is offset to the west in the mantle lithosphere. In other words, it implies that the mantle lithosphere continues to shorten beneath the Western Cordillera while the Eastern Cordillera grows above essentially undeforming mantle lithosphere. According to this scenario, the Central Andes behave as a single wedge including several back-thrusts, growing above a single mantle thrust located beneath the Western Cordillera. It requires a flat structure transferring shortening in the ductile crust over hundreds of kilometers, underneath the future Altiplano, with very little deformation, if any, in the overriding brittle crust. We contend that this scenario is mechanically untenable, because activating such a long flat structure requires a very low viscosity layer in the ductile crust to mechanically decouple the upper crust from the mantle lithosphere. This mechanism may activate in ultra-hot orogens, where the temperature of the lower crust has increased as a result of crustal thickening and high thermal gradients (e.g., Royden et al., 1997; Chardon et al., 2011), but not in the back-arc of the Central Andes before crustal shortening.

We propose instead that horizontal subduction triggered shortening in the hinterland of the mantle lithosphere rather than beneath the extinct Western Cordillera arc (Figure 6), resulting in the growth of a second complete crustal wedge overriding a mantle thrust zone, and that the width of the flat slab set the location of the Eastern Cordillera. Other processes, such as the reactivation of pre-Andean structures may also have influenced the precise location of thrust faults (e.g., Baby et al., 1995; Rochat et al., 1996; Sempere et al., 1997, 2002; Muller et al., 2002; Parks & McQuarrie, 2019), but we contend that slab flattening is the main cause that triggered the widening of the orogen.

The Eastern Cordillera that developed following slab flattening triggered the sudden eastward displacement of the foreland basin (e.g., deCelles & Horton, 2003; McQuarrie et al., 2005; Horton, 2005; Rak et al., 2017). Because the displacement of the retro-arc foreland basin depends on the distance between the two crustal wedges, and therefore on the length of the flat slab, it is not indicative of the amount of continental shortening. Foreland basin migration may only provide an estimate of the order of magnitude of the shortening where a single simple orogenic wedge is progressively growing and widening. Here, it only attests for the appearance of a second wedge, which implies that continental shortening cannot be quantified from the sole migration of the foreland basin (as considered for instance by deCelles & Horton, 2003).

**4.2.3 Tectonics of the Central Andes in the aftermath of the flat slab event**

Horizontal subduction was over by the end of the Oligocene in the Central Andes; the return to a more steeply dipping slab beneath the Bolivian orocline is evidenced by the widespread volcanic flare-up that affected the Eastern Cordillera, the Altiplano and the Western Cordillera area (Hoke et al., 1994; Sandeman et al., 1995; James & Sacks, 1999; O'Driscoll et al., 2012). This flare-up has been explained by the return of hot asthenosphere beneath the continental segment that had been hydrated but shielded from mantle heat by the horizontal slab (James & Sacks, 1999). This magmatic episode modified both the mechanical behavior of the continental plate and the boundary conditions that apply at the base of the crust (O'Driscoll et al., 2012). The strength of the hydrated and hotter continental lithosphere decreased, possibly facilitating the removal of dense mantle lithosphere that occurred at different periods underneath the Altiplano as well as beneath the Puna plateau (Garzione et al., 2006, 2017; Koulakov et al., 2006; Schurr et al., 2006; Leier et al., 2013; Heit et al., 2014; Currie et al., 2015; deCelles et al., 2015; Scire et al., 2016). The resulting new mechanical properties of the orogen rule out the analogy with the simple experiments simulating the growth of crustal wedges, such as those presented in Figures 1b and 2a, because these experiments assume that the mantle lithosphere is much stronger than the continental crust (see section 2.1 and Cagnard et al., 2006a). In the Neogene Central Andes, the hydrated and hotter lithosphere deforms easily, and the tectono-sedimentary evolution of the Altiplano attests for this behavior: Miocene shortening resulted in the development of a deep compressive hinterland basin in the Altiplano (e.g., Rak et al., 2017; Horton, 2018a), located between reverse faults of opposite vergence (Lamb and Hoke, 1997; Roperch et al., 1999; Perez et al., 2014). These basins may be compared to models simulating the shortening of a continental lithosphere in which the mantle is weak, where the geometry of crustal structures largely differs from crustal wedges developing above mantle megathrusts (Davy & Cobbold, 1991; Brun, 2002; Cagnard et al., 2006a, 2006b).

The rheology of the hydrated and hot lithosphere also favors the development of lower crustal flow from the over-thickened crustal wedges towards the thinner Altiplano, which may also contribute to thicken and uplift the plateau (Husson & Sempere, 2003; Gerbault et al., 2005; Eichelberger et al., 2015; Ouimet and Cook, 2010; Garzione et al., 2017). Low viscosity of the Altiplano middle and lower crust is supported by geophysical data : high electric conductivity (Brasse et al., 2002), low seismic velocity (Yuan et al., 2000; Ryan et al., 2016) both suggest partial melting in the middle crust beneath the high plateau. In fact, several mechanisms (lithosphere shortening, mantle removal, lower crustal flow) whose activation became possible following the end of the flat-slab event contributed to the major Neogene uplift of the Altiplano (see Garzione et al., 2017). Since the Pliocene, surface shortening in the Central Andes has focused in the lower Subandean ranges, possibly because buoyancy forces oppose further shortening of the most elevated parts of the range (Figure 6; e.g., Dalmayrac & Molnar, 1981; Froidevaux & Isacks, 1984; Sébrier et al., 1985).

**5. Discussion: from subduction dynamics to Andean tectonics, at different latitudes**

We expand this analysis of the Central Andean evolution in the Bolivian orocline to other latitudes. In this discussion, we first revisit the mechanisms that result in horizontal subductions, then review past and existing flat slab segments beneath South America and their effects on the Andes. We also review alternative mechanisms that have been proposed to explain the growth of this mountain range.

**5.1. Driving mechanisms for flat-slab subduction**

McGeary et al. (1985) noted that volcanic gaps are often associated with the presence of anomalously thick oceanic crust at trench. Indeed, the subduction of buoyant ridges and oceanic plateaux, whose effect is to make a slab segment neutrally or positively buoyant, favors slab flattening (e.g., Vogt, 1973; Vogt et al., 1976; Pilger, 1981; Gutscher et al., 1999; van Hunen et al., 2002a; Espurt et al., 2008; Antonijevic et al., 2015). Subducting buoyant plate segments, however, are not systematically associated with flat slabs (Skinner & Clayton, 2013). Models reproducing the effect of slab buoyancy on slab flattening show a delay between buoyant ridge subduction at trench and the appearance of a horizontal slab segment, which may partly explain this poor correlation (Espurt et al., 2008). Although analogue and numerical models confirm that the subduction of buoyant ridges favors slab flattening, Van Hunen et al. (2002a, 2002b) suggest that oceanic plateaus cannot produce flat slabs that extend further than 300 km inland, unless other mechanisms come into play, such as active overthrusting of the overriding plate toward the slab (see also Currie & Beaumont, 2011; Schepers et al., 2017). Other local features can promote flat-slab subduction, such as a thick continental lithosphere (O'Driscoll et al., 2009; 2012; Gerya et al., 2009; Hu et al., 2016) if it is also relatively cold (Rodriguez-Gonzalez et al., 2012; Manea et al., 2012; Taramon et al., 2015), a weak wedge on top of the slab (Manea and Gurnis, 2007; Sandiford et al., 2019), a buoyant sub-slab mantle material (Bishop et al., 2017), or slab break-off following the subduction of an oceanic plateau (Liu and Currie, 2016). Manea et al. (2017) reviews and compares some of these models.

Andean volcanism has been used to date slab flattening, and to show that flat slab segments different from those that exist today formed beneath South America during the Cenozoic (Figure 7, e.g., Ramos & Folguera, 2009). The slab entering the trench along South America has become progressively younger, and therefore less dense. The overriding motion of the continent toward the trench since the opening of the South Atlantic (Russo and Silver, 1996; Schellart et al., 2007; Husson et al., 2008; Martinod et al., 2010) facilitated the appearance of flat slab segments following the subduction of oceanic ridges (Gutscher et al., 1999; van Hunen et al., 2004). It explains why the horizontal subduction zone that developed beneath the Bolivian orocline in the Paleogene is prominent among a series of flat slab segments that appeared beneath South America at later times (Figure 7).

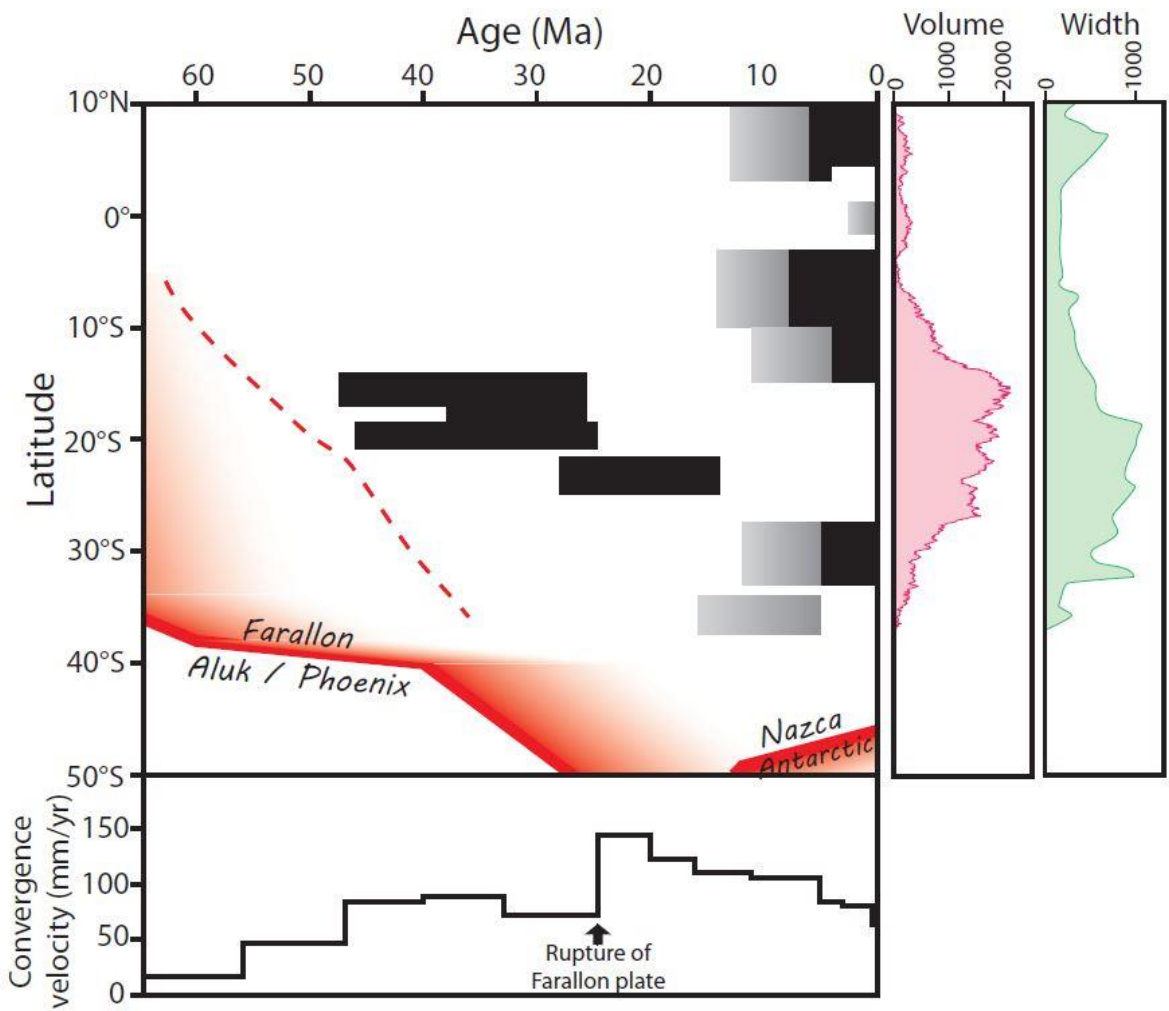

*Figure 7: Cenozoic horizontal slab segments beneath South America (after James & Sacks, 1999; Kay et al., 1999; Ramos et al., 2002; Bourdon et al., 2003; Rosenbaum et al., 2005; Ramos & Kay, 2006; Ramos & Folguera, 2009; Kay & Coira, 2009; Wagner et al., 2017). Gray intervals indicate periods of diminution of the slab angle marked by an eastward migration of the volcanic arc, and black intervals correspond to periods of horizontal subduction without any subduction-related volcanic activity. Red lines indicate oceanic ridge subduction at trench, after Chen et al. (2019) and Bretsprecher & Thorkelson (2009). The red dashed line marks the beginning of the Farallon slab anchoring in the lower mantle, according to Chen et al. (2019). The trench perpendicular subduction velocity at 17°S is reported after Somoza & Ghidella (2012). We also report the volume of the Andes above 2000 m (in km³/km) and the average width of the chain above 2000 m (in km).*

**5.2. Andean width and horizontal subduction zones**

At present-day, horizontal subduction zones can be imaged using seismic tomography and earthquake locations (e.g., Barazangi & Isacks, 1976; Cahill & Isacks, 1992; Whitman et al., 1996; Anderson et al., 2007; Alvarado et al., 2009; Kim et al, 2012; Mulcahy et al., 2014; Scire et al., 2015, 2016; Chiarabba et al., 2016; Syracuse et at., 2016; Lim et al., 2018). At the surface, flat-slab segments match the locations of gaps in the volcanic arc: between 27 and 33°S in Chile/Argentina, between 3 and 15°S in Peru (Barazangi & Isacks, 1976; Cahill & Isacks, 1992), and north of 6°N in northern Colombia (Chiarabba et al., 2016; Wagner et al., 2017) (Figure 7). Figure 7 also shows Cenozoic flat slab segments that developed beneath the Bolivian Orocline (Late Eocene-Oligocene, see above) and between 23 and 26°S (Miocene). We describe below for each ancient or current flat slab segment, the relationships existing between slab flattening and the eastward displacement of the location of continental shortening.

Continental deformation in Colombia shares similarities with the Paleogene Central Andes. The Eastern Cordillera in Colombia also is a bivergent thrust belt bounded by two marginal basins of opposite vergence. It results from the inversion of pre-existing Mesozoic rifting (Cooper et al., 1995; Mora et al., 2009, 2010). As in the Central Andes, the Eastern Cordillera appeared much later than the Western range (Central Cordillera) whose growth initiated in the Late Cretaceous (Colletta et al., 1990; Gomez et al., 2003, 2005; Bayona et al., 2008; Horton et al., 2010; Reyes-Harker et al., 2015). Slab flattening may have resulted from the subduction of a buoyant ridge that initiated 13 My ago (Wagner et al., 2017). Horizontal subduction, inferred from the cessation of volcanism north of 3°N, was achieved at 6 My, which implies that the delay between the onset of ridge subduction and slab flattening amounts to 7 My, akin to other segments of the Andes. Volcanism resumed 4 My ago between 3 and 5.5°N, evidencing a northward migration of the southern boundary of the Colombian flat-slab segment (Figure 7, Yarce et al., 2014; Wagner et al., 2017). The eastern end of the horizontal slab segment is located beneath the Eastern Cordillera and promotes its growth (Chiarabba et al., 2016; Siravo et al., 2019). Indeed, the growth of the Eastern Cordillera initiated timidly in the Paleogene, i.e. before the flat-slab episode (Bayona et al., 2008; Horton et al., 2010; Siravo et al., 2018), but the main episode of uplift of the range is late Neogene, concomitant with the appearance of the horizontal subduction, as attested both by tectonic, thermochronological and paleobotanical data (Gregory-Wodzicki, 2000; Mora et al., 2015; Siravo et al., 2019).

The Andes between 23 and 26°S are as high and wide as in the Bolivian Orocline. There, the Eastern Cordillera results from the inversion of the Salta Rift (Grier et al., 1991; Mon & Salfity, 1995; Cristallini et al., 1997; Kley et al., 2005; Carrera et al., 2006; Carrapa & deCelles, 2015). As elsewhere in the Andes, shortening and uplift of the forearc (Cordillera de Domeyko, west of the present-day Western Cordillera) initiated in the Late Cretaceous (Mpodozis et al., 2005; Arriagada et al., 2006; Carrapa & deCelles, 2015). Paleo-elevation analyses, sedimentological data combined with structural observations, suggest that the Eocene shortening has been sufficiently intense to thicken the crust and uplift the Western Cordillera and Western Puna Plateau at elevations approaching 4000 m as soon as during the Eocene (Carrapa & deCelles, 2015; Quade et al., 2015), although geomorphological observations in the forearc region show that the Western Cordillera further uplifted during the Neogene (Riquelme et al., 2003, 2007). Shortening and uplift of the Eastern Cordillera, in contrast, essentially began during the Middle-Late Miocene, i.e. much later than the growth of the Eastern Cordillera in the Bolivian oroclime (Carrera & Muñoz, 2008). Carrera et al. (2006) noted that former extensional faults dipping to the East and located West of the Lerma valley in the Eastern Cordillera were inverted. In contrast, West dipping faults were not inverted, suggesting again that the structure of the Eastern Cordillera is that of a bivergent crustal wedge. This has also been observed by Allmendinger & Gubbels (1996), who noted the westward vergence of the Santa Barbara Range compressive structures. At this latitude, this period coincides with a magmatic gap related to another flat-slab episode that developed between 22 and 25°S (Figures 4 and 7, Kay et al., 1994, 1999; Ramos & Folguera, 2009; Kay & Coira, 2009). We therefore propose that the Neogene episode of shortening in this segment of the Andes also corresponds to the growth of a crustal wedge located in the Eastern Cordillera that was triggered by horizontal subduction.

The Chile/Argentina flat slab segment (28°-33°S) may have been caused by the subduction of an oceanic passive ridge originated above the Juan Fernandez hot spot. Ridge subduction near 32°S began approximately 12 My ago (Yañez et al., 2001), resulting in the progressive eastward migration of the volcanic arc (Kay et al., 1999; Gutscher et al., 2000b; Ramos et al., 2002). The extinction of volcanism occurred 7 My later, indicating the amount of time required to flatten the slab (Espurt et al., 2008). Horizontal subduction is accompanied by an eastward migration of continental shortening, resulting in the uplift of the Sierras Pampeanas (Jordan et al., 1983; Ramos et al., 2002; Fosdick et al., 2015). Seismological data show that continental shortening concentrates above the extremity of the flat-slab segment (Pardo et al., 2002; Espurt et al., 2008). In contrast, the Western Cordillera that was deforming before slab flattening no longer accommodates any important shortening (Pardo et al., 2002; Rossel et al., 2016; Rodriguez et al., 2018).

In the Neuquen segment of the Southern Andes between 36° and 39°S, Kay et al. (2006) and Folguera et al. (2006) report Neogene variations in the dip of the slab suggested by the eastward progression of backarc magmatism and by changes in its geochemical characteristics. They note that the Early Miocene diminution of the angle of subduction correlated with the end of intra-arc extension (Charrier et al., 2005) and with the activation of thrust faults situated as far as the San Rafael block, more than 200 km East of the volcanic arc (Ramos et al., 2014). The analysis of volcanism suggests that the dip of the slab inclined again ~5 My ago in the Pliocene. This episode of slab shallowing corresponds to the Payenia flat subduction of Ramos & Folguera (2009). Nevertheless, the Andes at the latitude of the Neuquen Basin remain narrow, and shortening in the back-arc area has been moderate, probably because the slab never sufficiently flattened to become a horizontal subduction zone, as suggested by the permanence of magmatic activity along the arc throughout the Cenozoic (Kay et al., 2006). Therefore, interplate coupling was never fully achieved and did not trigger the growth of a second major crustal wedge further East at that latitude.

Considered together, these examples suggest that the appearance of a new eastern crustal orogenic wedge triggered by horizontal subduction is responsible for the widening of several segments of the Andes. Slab flattening was not coeval in the different regions described above explaining why the widening of the Andes was not concomitant in these segments.

However, the structural and deformation style of the range also varies largely from one Andean segment to another, and this cannot simply be explained by the diachronic appearance of flat slab segments. In the Central Andes for instance, surface shortening has been essentially concentrated in the thin-skinned structures of the Subandean ranges since 10 Ma, north of 23°S (Figure 3e), while thick-skinned tectonics has prevailed in the South (Allmendinger & Gubbels, 1996; McQuarrie, 2002b). Moreover, the Andes south of 23°S do not include any area accommodating minimum shortening between the two major wedges that may be compared to the Bolivian Altiplano. Several reasons may be invoked to explain these differences, such as the rheology of the lithosphere that accommodated the Andean orogeny. The pre-Andean rheology of the continent varied along-strike and partly controlled the growth of Andean structures. Geological observations indicate that the Altiplano and Puna have displayed structural differences for a long time, and Andean shortening could also vary according to preexisting heterogeneities, since there is a striking correlation between the change in deformation styles and the distribution of Paleozoic and Mesozoic basins (Martinez, 1980; Baby et al., 1989; Allmendinger & Gubbels, 1996; Kley et al., 1999; Coutand et al., 2001; McQuarrie, 2002b; Sempere et al., 2002). For instance, from ~21°S to 23°S, the dominantly Early Paleozoic sedimentary province in the eastern part of the orogen changes to a dominantly marine arc and younger Precambrian basement, which coincides with the termination of the thin-skinned Subandean belt. Then, the thinner mantle lithosphere observed in the Puna plateau compared to the Altiplano region (e.g., Whitman et al. 1996) may partly be inherited from Mesozoic extension (Kley, 1996; Sempere et al., 2002; Deeken et al., 2006; Insel et al., 2012). Thinner mantle lithosphere may then explain the different pattern of deformations south of 23°S compared to that described above in the Altiplano, and especially why the shortening has been more homogeneous in the Puna plateau (Allmendinger & Gubbels, 1996). Indeed, models in which the mantle lithosphere is not rigid do not result in the growth of crustal wedges localized above mantle thrusts but rather in wide domains of distributed shortening (Davy & Cobbold, 1991; Brun, 2002).

Hence, the comparison between structures at different latitudes show that the modes and distribution of deformation during Andean widening also depend on the mechanical properties of the continental plate which, in turn, depend on the pre-Andean history of the continent. Nevertheless, we emphasize that slab flattening, or absence thereof, is the primordial mechanism that sets the width of the Andes, at any latitude. Although the existence of distinct crustal wedges controlling the widening of the range is not as clearly visible in the absence of intra-mountainous poorly deformed segments comparable to the Altiplano, the number of west-verging crustal structures emerging in the eastern part of the range suggest they result from a wedge whose axis is located in the interior of the continent, eastward of the Western Cordilleran wedge.

## 5.3 Andean crustal wedges and vergence of lithospheric megathrusts

Mechanical models suggest that doubly vergent crustal wedges above lithospheric megathrusts should present an asymmetry resulting from the boundary conditions at the base of the crust (see section 2). Although major crustal faults with opposed vergence remain active on both sides of the wedge, shortening chiefly triggers the development of series of prograding thrusts in the external part of the wedge. On the other side of the range, the major back-thrust whose vergence is opposed to that of the mantle thrust remains the major structure that accommodates the growth of the wedge (Figure 1). Then, the evolution of deformation within the wedge gives insights into the vergence of deep structures.

The Andes at 33.5°S (latitude of Santiago de Chile) are a good example of doubly vergent crustal wedge with essentially west-vergent faults in Chile and east-vergent structures on the Argentinean side of the range. Armijo et al. (2010a) proposed that the San Ramon Fault that emerges East of Santiago is the main west-verging lithospheric-scale thrust that accommodates the growth of the range. East-verging faults emerging in Argentina would then correspond to back-thrusts located on the other side of the crustal wedge. This view has been contested by Astini & Davila (2010). These authors note that synorogenic strata in Argentina were progressively incorporated to the range while the thrust system migrated to the east during the Miocene, which suggests that the vergence of the deep mantle thrust underlying the crustal wedge is to the East (see Giambiagi et al., 2003; Horton, 2018b). The growth of the Andean wedge above an eastward verging major structure is also supported by Farias et al. (2010) based on seismological and structural data. In fact, the presence of a major crustal active west-verging structure located on the rear of a wedge growing above an east-verging mantle thrust is expected from the analysis of mechanical models (Malavieille, 1984; Naylor & Sinclair, 2008), and does not necessarily require that the vergence of the mantle lithosphere thrust is to the west.

Armijo et al. (2010b) note that the vergence of the wedge may change during the evolution of the orogen. Switching the vergence of the wedge implies changing the vergence of the thrust in the mantle lithosphere. This generally does not occur if the mantle lithosphere is the most rigid layer of the continental plate, as shown by numerical and analogue models (Martinod & Davy, 1994; Luth et al., 2010; Vogt et al., 2017; Jaquet et al., 2018). Some models suggest however that a change in the vergence of the deep mantle thrust may occur following slab break-off (e.g., Regard et al., 2008). More generally, if the mantle lithosphere is largely softened such that its strength becomes smaller than that of the continental crust, then the mechanical conditions of the convergence zone do not anymore correspond to those that are applied in crustal wedge models (Cagnard et al., 2006a; Burov & Watts, 2006). The mantle lithosphere may be sufficiently softened by advected fluids and heat in the magmatic arc area to permit changes in the vergence of the mantle thrust, although available surface data do not permit to evidence easily if such changes occurred in the Western Cordillera.

In the Bolivian orocline, in contrast, tectonic observations support a shift in the vergence of the thrust accommodating shortening in the mantle lithosphere beneath the Eastern Cordillera, from west to east, following the flat-slab episode (Figure 3). Indeed, deformation in the Eastern Cordillera essentially propagated westward towards the Altiplano between 43 and 27 Ma (Anderson et al., 2017; 2018). This evolution contrasts with the deformational regime that established since the end of the flat-slab episode with a thrust front advancing rapidly to the East. We propose that the hot and hydrated mantle lithosphere located above the Eocene-Oligocene flat slab segment permitted the inversion of the mantle thrust vergence. Geophysical studies, indeed, clearly confirm that the Brazilian craton is now underthrusting beneath the Andes (Isacks, 1988; Roeder & Chamberlain, 1995; Allmendinger & Gubbels, 1996; Allmendinger et al., 1997; Beck & Zandt, 2002; Oncken et al., 2012).

## 5.4 The effect of horizontal subduction on crustal rheology and its consequences on the modes of deformation, the elevation, and the volume of the Andes

Figure 7 presents the latitudinal distribution of the volume per unit length of the Cordillera above 2000 m. The volume of the Central Andes largely exceeds that of other sectors of the Cordillera, including where modern horizontal subduction is occurring, such as Colombia or Northern Peru. The particularity of the Central Andes is that episodes of flat-slab subduction occurred earlier than elsewhere, being achieved in the Late Oligocene–Miocene (Figure 7). It hydrated the continental plate, and the return to a steeply dipping slab resulted in the re-heating of this hydrated lithosphere, which resulted in a pulse of magmatism, including large volumes of ignimbritic flows (James & Sacks, 1999). Geophysical data confirm that partial melting is largely present (Brasse et al., 2002; Oncken et al., 2003; Ward et al., 2017) and suggest that the mantle lithosphere beneath the Altiplano and Puna has been partly removed (Garzione et al., 2006, 2014, 2017; Leier et al., 2013; Heit et al., 2014 Ryan et al., 2016).

Uplift may result from the removal of dense lower lithosphere (Bird, 1979, 1991; England & Houseman, 1988; Garzione et al., 2008, 2017). Efficient shortening of the hot and hydrated lithosphere favors continental shortening, resulting in further uplift (Lamb, 2011; Garzione et al., 2017). The reduced viscosity of the continental crust favors the development of both trench-perpendicular (Husson & Sempere, 2003) and trench-parallel (Ouimet and Cook, 2010; Gerbault et al., 2005) mid-lower crustal flows away from the zones of thick continental crust, which may explain the rather uniform elevation of the Altiplano, and why uplift also occurred in places where surface crustal shortening is insufficient to explain the observed crustal thickness (Figure 3e ; see e.g. Garzione et al. (2017) for discussion; England & Houseman (1988) for the description of mechanical processes). Crustal flow permits to relocate the mass excess from the eastern Andes, which may explain the Neogene uplift of the Western Cordillera that occurred without significant surface shortening (e.g., Farias et al., 2005; Thouret et al., 2007; Schildgen et al., 2007; Jordan et al., 2010). We emphasize that this mass transfer only became possible once the viscosity of the crust dropped due to hydration and heating, and should not have been active before and during horizontal subduction, when the continental crust was colder and thinner (see Figures 3d and 3e).

Overall, the thermo-mechanical conditions that were established following the end of the flat-slab episode may explain the Neogene uplift of the Altiplano region. The average Andean elevation remains modest above present-day horizontal slab segments, even though the width of the chain has already increased there (Figure 7). We conclude that the bulk elevation increase largely occurs after the slab returns to a steeper dip.

## 5.5 Other geodynamical parameters involved in the growth and widening of the Andes

At least two geodynamical mechanisms other than slab flattening have been proposed to explain the Andean segmentation and the appearance of the higher and wider Central Andes. They are driven by (i) toroidal mantle flow that accommodates the westward retreat of the slab (e.g., Russo & Silver,1994; Silver et al., 1998; Schellart et al., 2007), and (ii) slab anchoring in the lower mantle (e.g., Quinteros & Sobolev, 2013; Faccenna et al., 2017; Chen et al., 2019). In this section, we discuss these two mechanisms and analyze how they may contribute to variations in the timing and amplitude of the Andean growth.

(i) The advance of the overriding plate towards the trench requires slab retreat in the lower mantle reference frame. The mantle must flow to accommodate the retreating motion of the slab (Garfunkel et al., 1986; Russo & Silver, 1994; Olbertz et al., 1997; Schellart, 2004, 2008; Funiciello et al., 2006; Guillaume et al., 2010). When the slab reaches the lower mantle, toroidal displacements increase in the upper mantle: the mantle laterally bypasses the slab to accommodate its roll back (Funiciello et al., 2006). Russo and Silver (1994) noted that the Central Andes are located far from the lateral edges of the slab and proposed that the larger amount of continental shortening accommodated in the Central Andes resulted from the effect of toroidal mantle flows accommodating slab roll back. Slab retreat there is hampered because more energy is needed to achieve the required toroidal flow (see also Schellart et al., 2007; Armijo et al., 2015). This process may in turn explain the additional shortening occurring in this part of the Andes, at the center of the present-day subduction zone. However, we note that rapid shortening in the Central Andes began in the Eocene, at a time when the Farallon Plate that subducted beneath South America was much larger than the present-day Nazca plate (Figure 8). The northern edge of the plate was located beneath North America, while the southern one was located beneath Patagonia (Wright et al., 2016). If the distance to the slab edges was the prominent control on shortening in South America, the main episode of shortening should have had a larger extent and be centered in Northern Peru, which was even farther from the slab boundaries than the growing Bolivian orocline.

(ii) Interactions between the slab and the lower mantle beneath South America have also been proposed to control the subduction kinematics (e.g., Quinteros & Sobolev, 2013) and in turn, the growth of the Andes (Faccenna et al., 2017;

Chen et al., 2019). Westward slab retreat is hindered by slab anchoring in the lower mantle. Based on tomographic data and plate reconstructions, Faccenna et al. (2017) and Chen et al. (2019) note that the subduction of the Farallon plate beneath South America is much older in its northern half (Figure 7). Another independent piece of evidence in favor of a younger Farallon subduction beneath the Southern Andes is given by plate kinematics reconstructions. Somoza & Ghidella (2012) note that the position of the Euler pole that describes the clockwise rotation of Farallon with respect to South America was located in Brazil in the Upper Cretaceous and Paleocene. They conclude that the two plates were diverging at the latitudes of the Southern Andes, and that another oceanic plate, namely the Aluk (or Phoenix) plate was subducting beneath that part of South America (see also Wright et al., 2016). Chen et al. (2019) propose that the anchoring of the Farallon slab occurred beneath the Northern and Central Andes during the Paleocene (Figure 7), which may explain the Eocene pulse of shortening that resulted in the growth of the Central Andes. This mechanism may also explain the extension that occurred simultaneously in the southern Andes, and why the onset of shortening occurred later to the south. Indeed, models confirm that the effect of slab pull triggering continental extension is larger before slab anchoring in the lower mantle (Faccenna et al., 2001; Quinteros & Sobolev 2013). Older slab anchoring beneath Northern South America does not explain, however, why crustal shortening was much more intense in the Bolivian orocline than in Northern Peru (Kley & Monaldi, 1998; Oncken et al., 2006; Gotberg et al., 2010; Perez et al., 2016; Garzione et al., 2017; Faccenna et al., 2017), as evidenced by the variations of Andean volume vs. latitude (Figure 7).

Although both the effect of slab anchoring and toroidal mantle flow may have contributed to the growth and segmentation of the Andean Range, they do not clearly explain why the Bolivian orocline is higher and larger than the northern Peruvian Andes, for instance. We propose that their effects, combined with those resulting from slab flattening, explain the different latitudinal chronology in the widening of the Andean range.

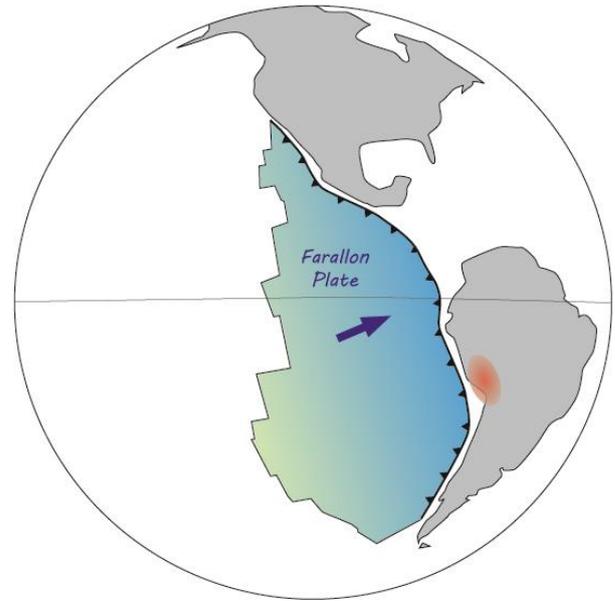

*Figure 8:* Farallon Plate subducting beneath America at the Eocene-Oligocene boundary (33 My ago), after Wright et al. (2016). The positions of Southern Peru and Northern Chile with respect to the lateral boundaries of the slab do not explain the emergence of the Bolivian orocline at that time. The red area underlines the position of the Late Eocene horizontal slab segment.

**Conclusion**

We argue that the occurrence of horizontal slab segments favored the hinterland migration of the shortening in South America. Models and observation of major orogens show that tectonic structures that accommodate continental shortening generally correspond to crustal wedges growing above mantle lithosphere thrusts. In the Andes, the appearance of flat-slab segments triggers the growth of crustal wedges in the continental hinterland, distinct from the crustal wedge that grows closer to the trench when the slab is steeply dipping. In the Bolivian orocline, the Eastern Cordillera is the most mature example of a doubly-vergent crustal wedge that appeared hundreds of kilometers to the East of the earlier Western Cordilleran wedge, following the Late Eocene-Oligocene episode of slab flattening. The same mode of hinterland migration of shortening also occurred later in the Neogene at different latitudes as in the Northern (Colombia) and Central Andes (NW Argentina). Different mechanisms activate once the strength of the continental plate decreases, following the end of flat-slab subduction, which explains the delay between the growth of crustal wedges and the appearance of the high plateau (Altiplano). We conclude that horizontal subduction has a major effect the geometry and evolution of Andean crustal structures, and largely explains the episodic and geographically segmented evolution of this range.

**Acknowledgments:** The authors warmly thank the editor Douwe Van Hinsbergen, Nadine McQuarrie and another anonymous reviewer for their very detailed and constructive comments. We also thank Marcelo Farias and Pierrick Roperch for sharing with us their knowledge of the Andes.